
\input phyzzx.tex
\pubnum{ROM2F-94-44}

\def\b{\beta}

\def\m{\mu}
\def\n{\nu}

\def\s{\sigma}
\def\t{\tau}

\def\mnr{\mu\nu\rho}
\def\mn{\mu\nu}

\def\wh{\widehat}
\def\wt{\widetilde}
\def\de{\partial}

\def\eps{\varepsilon^{\alpha\beta}}

\def\mezzo{{1 \over 2}}
\def\unoar{1-({a \over r})^4}

\def\ap{\alpha^\prime}
\def\unita{{1 \kern-.30em 1}}
\def\fey{{\big / \kern-.80em D}}
\def\complex{{\kern .1em {\raise .47ex \hbox {$\scriptscriptstyle |$}}
\kern -.4em {\rm C}}}
\def\zet{{Z \kern-.45em Z}}
\def\real{{\vrule height 1.6ex width 0.05em depth 0ex
\kern -0.06em {\rm R}}}
\def\rational{{\kern .1em {\raise .47ex \hbox{$\scripscriptstyle |$}}
\kern -.35em {\rm Q}}}
\titlepage
\title{String Theory and ALE Instantons\foot{Talk given by F.Fucito
at the XI SIGRAV meeting, Trieste, September 27-30, 1994.}}
\author{M.~Bianchi, F.~Fucito, G.C.~Rossi}
\address{Dipartimento di Fisica, Universit\`a di Roma II ``Tor
Vergata''
I.N.F.N. Sezione di Roma II ``Tor Vergata'',00133 Roma, Italy}
\andauthor{ M.~Martellini}
\address{Dipartimento di Fisica, Universit\`a di Milano, 20133
Milano, Italy
Sezione I.N.F.N. dell'Universit\`a di Pavia, 27100 Pavia, Italy}

\abstract
{We show that the classical equations of motion of the low-energy effective
field theory describing the massless modes of the heterotic (or type I)
string admit two classes of supersymmetric self--dual backgrounds.
The first class, which was already considered in the literature,
consists of solutions with a (conformally) flat metric coupled to
axionic instantons.
The second includes Asymptotically Locally Euclidean (ALE) gravitational
instantonic backgrounds coupled to gauge instantons
through the so--called  ``standard embedding''.}
\endpage
\pagenumber=1

\REF\eh{T.~Eguchi and A.J.~Hanson, Ann.Phys.{\bf 120} (1979) 82.}
\REF\egh{For a review on gravitational instantons see: T.~Eguchi,
P.B.~Gilkey and A.J.~Hanson, Phys.Rep. {\bf 66} (1980) 213.}
\REF\rossi{V.~Novikov, M.~Shifman, A.~Vainshtein and
V.~Zakharov, Nucl.Phys. {\bf B260} (1985) 157; \nextline
D.~Amati, G.C.~Rossi and G.~Veneziano, Nucl.Phys. {\bf B249} (1985) 1;\nextline
I.~Affleck, M.~Dine and N.~Seiberg, Nucl.Phys. {\bf B256} (1985) 557;\nextline
D.~Amati, K.~Konishi, Y.~Meurice, G.C.~Rossi and
G.~Veneziano, Phys.Rep. {\bf 162} (1988) 169.}
\REF\wittuno{E.~Witten, Nucl.Phys. {\bf B185} (1981) 513.}
\REF\konuno{K.~Konishi, N.~Magnoli and
H.~Panagopoulos, Nucl.Phys. {\bf B309} (1988) 201; ibid.
{\bf B323} (1989) 441.}
\REF\hp{S.W.~Hawking and
C.N.~Pope, Nucl.Phys. {\bf B146} (1978) 381.}
\REF\noi{M.~Bianchi, F.~Fucito, M.~Martellini and G.C.~Rossi, in preparation.}
\REF\kontre{K.~Konishi, Phys.Lett. {\bf B135} (1984) 439.}
\REF\peter{P.K.~Townsend and
P.~van Niewenhuizen, Phys.Rev.{\bf 19D} (1979) 3592.}
\REF\taylor{S.-J.~Rey and T.R.~Taylor, Phys.Rev.Lett. {\bf 71} (1993) 1132.}
\REF\rey{S.-J.~Rey, Phys.Rev. {\bf D43} (1991) 526.}
\REF\calluno{C.~Callan, J.~Harvey and
A.~Strominger, Nucl.Phys {\bf B359}(1991) 611; ibid. {\bf B367} (1991) 60.}
\REF\khuri{R.R.~Khuri, Nucl.Phys. {\bf B387} (1992)
315.}
\REF\torinesi{D.~Anselmi,
M.~Billo', P.~Fre', L.~Girardello and A.~Zaffaroni, {\it ALE Manifolds and
Conformal Field Theories}, SISSA/44/92/EP, hep-th/9304135}
\REF\sigmaquattro{L.~Alvarez-Gaum\'e and D.Z.~Freedman, Phys.Rev.
{\bf D15} (1980) 846; Comm. Math. Phys. {\bf 80} (1981) 443;\nextline
L.~Alvarez-Gaum\'e, Nucl.Phys. {\bf 184} (1981) 180;\nextline
L.~Alvarez-Gaum\'e and P.~Ginsparg, Comm. Math. Phys. {\bf 102}
(1985) 311;\nextline
C.~Hull, Nucl.Phys. {\bf 260} (1985) 182.}
\REF\antoniadis{I.~Antoniadis, E.~Gava, K.~Narain and T.~Taylor,
{\it Topological Amplitudes in String Theory}, ICTP preprint
IC/93/202;\nextline
E.~Witten, Comm.Math.Phys. {\bf 117}(1988)353.}

Asymptotically Locally Euclidean (ALE) Instantons have played a
fundamental role in Euclidean Quantum (Super)-Gravity [\eh, \egh].
Furthermore, it is hoped that they may be responsible for
the formation of a gravitino condensate which may trigger
dynamical supersymmetry (SUSY) breaking [\wittuno, \konuno, \hp].

In this talk, after reviewing the role of fermionic condensates in
the problem of  SUSY breaking, we will discuss how to
promote gravitational ALE instantons to full-fledged
solutions of the heterotic string equations of motion
and we will study the geometrical properties of the simplest among them,
an Eguchi-Hanson instanton coupled to a gauge
instanton through the ``standard embedding".
This investigation is propaedeutical to a saddle-point evaluation of
instanton dominated Green functions in the low-energy effective
supergravity arising from the heterotic string compactified to D=4 [\noi].

Let us first turn to a brief discussion of the role of fermionic condensates
in SUSY breaking. As is well-known, a globally SUSY theory has SUSY vacua
with zero energy. The Hamiltonian of the theory can in fact be written as a
quadratic form of the SUSY charges, $Q$.
A signal of SUSY breaking is thus provided by a non zero
vacuum expectation value (v.e.v.)
$$
\bra{0}\delta X\ket{0}=\bra{0}\{Q,X\}\ket{0}\neq 0
\eqn\susyuno
$$
that can be interpreted as the charge $Q$ not annihilating the vacuum.

Let us introduce chiral and vector superfields(in the Wess-Zumino gauge)
$$
\cases{
\Phi=z+\sqrt2\theta\chi+\theta^2 F\cr
V=-\theta\sigma_\mu\bar\theta A^\mu+i\theta^2\bar\theta\bar\lambda
-i\bar\theta^2\theta\lambda+\mezzo\theta^2\bar\theta^2 D\cr}
\eqn\susydue
$$
Under a SUSY transformation with parameter $\epsilon$, the fermions
transform as
$$
\cases{
\delta\chi=\sqrt2\epsilon F+i\sqrt2\bar\epsilon\sigma^\mu\partial_\mu z\cr
\delta\lambda=\epsilon\sigma^{\mu\nu}F_{\mu\nu}+\epsilon D\cr}
\eqn\susytre
$$
Using \susytre~we find
$$
\bra{0}\delta \psi\ket{0}\approx \bra{0}\{Q,\psi\}\ket{0},\quad\quad
\bra{0}\delta \lambda\ket{0}\approx\bra{0}\{Q,\lambda\}\ket{0}
\eqn\susyquattro
$$
as the v.e.v's of all the other fields (and of their derivatives)
contained in \susytre~are supposed to vanish. SUSY is thus broken if and only
if the auxiliary fields get a non zero v.e.v. The fermionic partner
of the auxiliary field that develops a v.e.v. is the massless Goldstone
fermion (goldstino)
of broken SUSY. The scalar potential of the theory is
$$
V=FF^*+\mezzo D^2
$$
and a non-vanishing v.e.v for the auxiliary fields implies a vacuum energy
different from zero in agreement with the form of the Hamiltonian of a SUSY
theory.

For a local SUSY theory the matter is a little more complicated. The auxiliary
fields which are the partners of spin $\mezzo$ chiral and gauge fermions are
$$
\eqalign{
F_i&=e^{-{G\over 2}}(G^{-1})^j_iG_j+{1\over 4}f_{ab k}(G^{-1})^k_i
\lambda^a\lambda^b\cr
&-(G^{-1})^k_iG^{jl}_k\chi_j\chi_l-\mezzo\chi_i(G_j\chi^j),\cr
D_a&=i \Re f^{-1}_{ab}(-qG^iT^{b j}_i z_j
+\mezzo if^i_{bc}\chi_i\lambda^c\cr
&-\mezzo if_i^{*bc}\chi^i\lambda_c)
-\mezzo\lambda_a(G^i\chi_i)\cr}
\eqn\susycinque
$$
In the above formula $(z_i,\chi_i)$ are the components of a chiral matter
multiplet and $i$ is a gauge group index, $T^{\alpha j}_i$ is the group
generator and $q$ the gauge coupling constant.
 $f^i_{\alpha\beta}={\partial f_{\alpha\beta}\over\partial z_i},
G^i={\partial G\over\partial z_i}$ where $f$ is a coupling function
depending on the
chiral field $z$and $G$ is the K\"ahler potential.
{}From \susycinque~we see that the auxiliary fields develop a v.e.v. if the
fermion bilinears do. This should convince the reader of the importance of the
study of fermionic correlators in order to see whether or not
they can develop a v.e.v. when non-perturbative instanton effects are
taken into account.

In the case of globally SUSY theories a detailed study of a possible
breaking mechanism of SUSY via gauge instantons was carried out in [\rossi]
with rather encouraging results. The case of locally supersymmetric theories
was addressed to in [\hp, \wittuno], where it was argued that, in
supergravity, gravitino condensation due to gravitational instantons
can trigger the breaking of SUSY. An important step forward in this line
of arguments was made in [\konuno], where,
in the case of $N=1$ supergravity, the gravitino field strength condensate,
$<\psi_{ab} \psi^{ab}>$, was indeed computed and shown to be finite and
space--time independent. The
calculation was done by performing a saddle--point approximation around the
non trivial classical solution of
the theory represented by the Eguchi--Hanson gravitational instanton [\egh].

The idea behind this kind of approach is the hope of being able to infer
SUSY breaking by putting non trivial condensates in relation with some
anomalous SUSY transformation.
In globally SUSY theories in flat space [\rossi] such a relation exists and
it is called the Konishi anomaly equation [\kontre]. For a Super Yang-Mills
theory coupled to chiral scalar matter
multiplets, $\Phi^i$, belonging to the representations $\underline R_{\>}^i$
of the gauge group, the relevant anomalous commutator reads
$$
\{\bar Q,\bar\chi_i z^j \} = {q^2 c_i\over 32 \pi^2} \lambda\lambda
\delta_i^j+z^j{d \bar W \over d \bar z_i}
\eqn\konsei
$$
where $i,j$ are flavour indices, $W$ is the superpotential,
$\bar Q$ is a SUSY charge, $q$ the gauge coupling constant,
$\lambda$ is the gaugino, $z^i$ and $\chi^i$ are components of the chiral
multiplet, $\Phi^i$,  and $c_i$ is the index of the representation
$\underline R_i$.
The anomaly equation \konsei~lies in the same supermultiplet as the chiral
anomaly,
\ie~it is a partner of the anomalous divergence equation of the $R$-symmetry
current of the theory. The discovery that (some of) the condensates appearing
in \konsei~acquire a non-zero vacuum expectation value allowed
to derive detailed information on the degeneracy
pattern of the vacuum state manifold and, in some cases (SUSY theories with
no flat directions and chiral matter in suitably choosen representations
of the gauge group) even to conclude that SUSY is
dynamically broken by non--perturbative instanton effects.

To construct a similar argument in supergravity
one has to start with the anomalous divergence of the $R$-symmetry chiral
current, $j^\mu$
$$
D_\mu j^\mu=-{1 \over 384 \pi^2} R_{abcd}\wt R^{abcd}
\eqn\konanomalia
$$
where the tilde stands for the duality operation. Since
$R_{abcd}\wt R^{abcd}$ is the top component of a chiral multiplet,
which has $\psi_{ab} \psi^{ab}$ as lowest component [\peter], the analogue of
the anomalous SUSY transformation \konsei~is
$$
\{\bar Q,\bar\lambda \phi\}={ \kappa^2 \over 384\pi^2}\psi_{ab} \psi^{ab}
\eqn\basta
$$
where $\kappa$ is the gravitational coupling constant, $\psi^{ab}$ is the
gravitino field--strength and $\lambda$ and $\phi$ are components of a chiral
matter multiplet (for the sake of definiteness they may be respectively taken
to be
the dilatino and the dilaton). The appearance of the gravitino field--strength
bilinear in the right--hand--side of \basta~led correctly the authors
of [\konuno] to compute the expectation value of $\psi_{ab} \psi^{ab}$,
instead of $\psi_{\m} \psi^{\m}$, as suggested in [\hp].

Many of the computations we will present in this paper are very much in
the same line of thought of instanton calculus in globally SUSY
theories [\rossi], where, as we said, the simultaneous presence
of globally SUSY and of classical instantonic solutions conspire to
give exact space--time independent constant results for certain correlators.
The calculations we present here
are propaedeutical to extend instanton calculus to locally supersymmetric
(\ie~to supergravity) theories [\taylor, \noi].

Contrary to globally supersymmetric
Yang-Mills theories, however, supergravity is not renormalizable.
Strictly speaking, this puts the entire subject of instanton calculus in
supergravity on a rather shaky basis. On this problem we would like to take
the point of view (as also
suggested in [\konuno]) that supergravity theories should be considered
as low energy limits of string theories, which are expected not to suffer
from these deficencies. Thus to the order to
which supergravity theories are formally renormalizable (i.e. generically up to
two loops) results from perturbative and non--perturbative (instanton)
calculations should be considered as the limiting values of the corresponding
exact string results.

There are two more reasons to pursue this philosophy we would like to
mention here. The first has to do with the fact that effective
field theories appear at the moment the only arena where non--perturbative
aspects of string theories can be studied. The second is that,
exploiting the relation between bosonic zero--modes and instanton
``collective coordinates'', explicit computations of the former
may prove to be a useful starting point in the investigation of the structure
of the istanton moduli space over non-compact ALE manifolds.
Except for the purely gravitational sector, this subject is as yet poorly
understood.

We now give a brief outline of our results, referring the interest reader
to our original paper [\noi] for more details.

Non trivial supersymmetric solutions of the lowest order (in $\ap$) equations
of motion may be found setting to zero the fermion fields together with their
supersymmetric variations. A supersymmetric ansatz for the solution
is given by [\rey, \calluno, \khuri]:
$$
F_{\mu\nu} = \wt F_{\mn} \quad
H_{\mnr}= \sqrt{G} \varepsilon_{\mnr}{}^\sigma\de_\sigma\phi \quad
G_{\mn} =  e^{2\phi} \wh g_{\mn} \quad
\eqn\ansatz
$$
where $F_{\mn}$ is the field-strength of the gauge fields $A_\m$, $H_{\mnr}$
is the (modified) field-strength of the antisymmetric tensor $B_{\mn}$
and $\wh g_{\mn}$ is a self-dual metric. The generalized spin-connection
with torsion $\Omega_{\m\pm}^{ab}=\omega_\mu^{ab}\pm H^{ab}_\mu$
deriving from \ansatz~ is self-dual.
\ansatz~ must be supplemented with the Bianchi identity:
$dH=\ap \{ trR(\Omega_{-})\wedge R(\Omega_{-})-trF(A)\wedge F(A) \}$.
To simplify matters it is convenient to impose the ``standard embedding''
of the gauge connection in the $SU(2)$ spin group: $A=\Omega_-$.
There are two options. The first has been considered in [\rey,
\calluno, \rey] and leads to
conformally flat axionic instantons.
We would like to concentrate on the other, \ie~
$ A^i_\mu ={1 \over 2} \eta^i_{ab} \Omega_{\m-}^{ab}
= {1 \over 2} \eta^i_{ab} \wh\omega_{\m}^{ab}$, whose consistency requires
a constant dilaton and a vanishing torsion [\noi].
In this case, \ansatz~is completely specified by the choice of a self-dual
metric $\wh g_{\mn}$.

An interesting class of self-dual metrics is given by the
Gibbons-\break
Hawking multi-center (GHMC) ansatz [see, \eg \egh]:
$$
ds^2 = V^{-1}({\vec x}) (d\t + {\vec\omega}\cdot d {\vec x})^2 +
V({\vec x})d{\vec x}\cdot d {\vec x}
\eqn\multicenter
$$
with ${\vec \nabla}V={\vec \nabla}\times{\vec\omega}$ and
$V({\vec x}) = \eps + 2m \sum_{i=1}^{ k+1} {1\over\mid {\vec x}-
{\vec x}_i\mid}$.
The choice $\eps=0, m=\mezzo$ corresponds to ALE metrics.
ALE manifolds are smooth resolutions of singular varieties
in $\complex^3$ and are completly classified in terms
of the kleinian subgroups $\Gamma$ of $SU(2)$ [see, \eg \egh].
ALE instantons are non-compact Ricci-flat hyperk\"ahler manifolds of
$SU(2)/\Gamma$ holonomy and deserve to be considered
as non-compact Calabi-Yau manifold of complex dimension two [\torinesi].

The non-linear $\s$-model describing the propagation of the
heterotic string on ALE instantons with the standard embedding is left-right
symmetric and admits $N=(4,4)$ supersymmetry [\sigmaquattro].
The corresponding $\b$-functions vanish to all orders [\sigmaquattro] thus
the lowest order ALE solutions receive no radiative corrections in $\ap$.
Indeed, at the singular point of the moduli space where ALE instantons
coincide with algebraic varieties, string propagation is governed by a
$\complex^2/\Gamma$ orbifold conformal field theory [\torinesi].

We now turn to describe some geometrical properties of the
heterotic solution based on the EH (Eguchi-Hanson) instanton [see, \eg \egh]:
$$
ds^2= \wh g_{\mn} dx^\mu dx^\nu=
({r\over u})^2 dr^2+r^2 (\sigma_x^2+\sigma^2_y)+ u^2 \sigma_z^2
\eqn\ehmetric
$$
To remove the bolt singularity at $r=a$, we change the radial variable to
$u=r \sqrt{\unoar}$ and identify antipodal points. The EH instanton
has an $S^3/Z_2$ boundary and admits
an $SU(2)_R \otimes U(1)_L$ isometry group. Its Euler characteristic $\chi$
is two and its Hirzebruch signature $\tau= b^+_2-b^-_2$ is one.
Exploiting the global right-handed supersymmetry generated by the
covariantly constant spinor $\bar\epsilon$,
the two (left-handed) gravitino zero-modes on EH may be
expressed in terms of the closed self-dual two-form [\hp, \konuno].
These zero-modes are generated by the broken global left-handed
supersymmetry with parameter $\eta=u\eta_0$ ($\eta_0=i\s_0\bar\epsilon$):
$\psi_\mu=D_\mu \eta-{1\over 4}\s_\m \> \fey\eta$ [\noi].
The three zero-modes of the metric can be obtained
performing a further supersymmetry transformation:
$h_{\mn}= \nabla_\m \xi_\n + \nabla_\n \xi_\m - \mezzo (\nabla\cdot\xi)g_{\mn}$
with $\xi_\m = \eta\s_\m\bar\epsilon$.
A suitable interpretation of the infinitesimal diffeomorphisms $\xi^\m$
allows to relate them to the lack of invariance of the EH metric
under dilatation and the two rotations in the coset $SU(2)_L/U(1)_L$.
No dilatino zero-modes are expected on EH since the index of the Dirac
operator (for gauge singlets) is zero [see, \eg \egh].
However the presence of ``charged" spinor zero-modes is guaranteed by
a non-vanishing value of the index of the Dirac operator coupled to
the gauge bundle $V$.
Embedding the instantonic $SU(2)$ in the ``hidden" $E(8)$ only the gauginos
will be affected.  With respect to the subgroup, $SU(2)\otimes E(7)$,
the adjoint of $E(8)$ breaks as: $\underline{248}=(\underline3,\underline1)
\oplus(\underline2,\underline{56})\oplus (\underline1,\underline{133})$.
The general formula for the index of the Dirac operator, $\fey_V$, coupled
to a vector bundle $V$ may be found in [see, \eg \egh]. Performing the
necessary
manipulations, we find:
$ind(\fey_{\underline2},M,\de M)=-1=\tau$
and $ind(\fey_{\underline3},M,\de M)=-6$.
Since $\fey_V$ has no right-handed zero-modes, one (normalizable)
left-handed gaugino zero-mode is expected for each of the 56 doublets
and six for the triplet.
The explicit expression of the gaugino zero-modes in the doublet can be easily
found by exploiting the explicit form of the normalizable harmonic self-dual
two-form on the EH manifold.
The six triplet zero-modes may be found by explicitly solving the Dirac
equation, but we haven't been able to provide a geometrical interpretation
[\noi].
Thanks to the global right-handed supersymmetry, each zero-mode of the
gaugino generates two zero-modes for the gauge fields.

As is well known the bosonic zero-modes are related to the collective
coordinate of the heterotic EH background.
Thanks to the cancellation of the non-zero-mode functional determinants,
the evaluation of the relevant correlation
functions reduces to an integration of zero-modes of fermi fields
over the finite-dimensional moduli space of the heterotic EH instanton.
In [\noi] the saddle-point approximation is performed and the consistency
of the result with Ward identities of a properly defined global supersymmetry
is checked. The interpretation in terms of condensates as well as
the relation to topological amplitudes in string theory and
to topological field theories [\antoniadis] is under investigation [\noi].
\endpage
\refout
\end